\documentclass[10pt]{article}

\usepackage{amsmath,amssymb,amsthm}
\usepackage{amsxtra}
\usepackage{amsfonts}
\usepackage{amssymb}
\usepackage{url}
\usepackage{hyperref}
\usepackage{esint}
\usepackage{eucal}
\usepackage{mathrsfs}
\usepackage{cite}
\usepackage{graphicx,graphics}
\usepackage{stackengine}
\setstackgap{S}{1pt}
\newcommand\fig[1] {{\rm Figure}~\ref{fig:#1}}
\newcommand\labfig[1] {\label{fig:#1}}

\newcommand\labsect[1] {\label{sect:#1}}

\newcommand\eq[1] {(\ref{#1})}
\newcommand{\bfm}[1]{\mbox{\boldmath ${#1}$}}

\newcommand{\nonum}{\nonumber \\}
\newcommand{\beqa}{\begin{eqnarray}}
\newcommand{\eeqa}[1]{\label{#1}\end{eqnarray}}
\newcommand{\beq}{\begin{equation}}
\newcommand{\eeq}[1]{\label{#1}\end{equation}}
\newcommand{\Grad}{\nabla}
\newcommand{\Div}{\nabla \cdot}
\newcommand{\Curl}{\nabla \times}

\newcommand{\Tr}{\mathop{\rm Tr}\nolimits}

\newcommand{\Md}{\partial}

\newcommand{\Ga}{\alpha}
\newcommand{\Gb}{\beta}

\newcommand{\Ge}{\epsilon}
\newcommand{\Gve}{\varepsilon}

\newcommand{\Gk}{\kappa}

\newcommand{\Gl}{\lambda}
\newcommand{\Gn}{\eta}
\newcommand{\Gm}{\mu}

\newcommand{\Gr}{\rho}

\newcommand{\Gs}{\sigma}

\newcommand{\Go}{\omega}

\newcommand{\Gy}{\psi}

\newcommand{\BGe}{\bfm\epsilon}

\newcommand{\BGn}{\bfm\eta}
\newcommand{\BGm}{\bfm\mu}

\newcommand{\BGt}{\bfm\theta}

\newcommand{\BGr}{\bfm\rho}

\newcommand{\BGs}{\bfm\sigma}

\newcommand{\BGy}{\bfm\psi}

\newcommand{\BGG}{\bfm\Gamma}
\newcommand{\BGL}{\bfm\Lambda}

\newcommand{\CD}{{\cal D}}
\newcommand{\CE}{{\cal E}}

\newcommand{\CJ}{{\cal J}}

\newcommand{\CT}{{\cal T}}

\newcommand{\BCA}{{\bfm{\cal A}}}

\newcommand{\BCC}{{\bfm{\cal C}}}
\newcommand{\BCD}{{\bfm{\cal D}}}

\newcommand{\bpm}{\begin{pmatrix}}
\newcommand{\epm}{\end{pmatrix}}

\def\Bb{{\bf b}}

\def\Bd{{\bf d}}
\def\Be{{\bf e}}
\def\Bf{{\bf f}}

\def\Bh{{\bf h}}

\def\Bj{{\bf j}}
\def\Bk{{\bf k}}

\def\Bm{{\bf m}}
\def\Bn{{\bf n}}

\def\Bp{{\bf p}}
\def\Bq{{\bf q}}
\def\Br{{\bf r}}
\def\Bs{{\bf s}}
\def\Bt{{\bf t}}
\def\Bu{{\bf u}}
\def\Bv{{\bf v}}

\def\Bx{{\bf x}}

\def\BA{{\bf A}}

\def\BC{{\bf C}}
\def\BD{{\bf D}}
\def\BE{{\bf E}}

\def\BG{{\bf G}}
\def\BH{{\bf H}}
\def\BI{{\bf I}}
\def\BJ{{\bf J}}

\def\BL{{\bf L}}
\def\BM{{\bf M}}

\def\BP{{\bf P}}

\def\BR{{\bf R}}
\def\BS{{\bf S}}
\def\BT{{\bf T}}
\def\BU{{\bf U}}
\def\BV{{\bf V}}

\usepackage{graphics}
\usepackage{amssymb}

\title{A unifying perspective on linear continuum equations prevalent in science. Part IV: Canonical forms for equations involving higher order gradients}
\author{}
\date{}
\begin{document}
\maketitle
\vskip -.5cm
\centerline{\large Graeme W. Milton}
\centerline{Department of Mathematics, University of Utah, USA -- milton@math.utah.edu.}
\vskip 1.cm
\begin{abstract}
  Enlarging on Parts I, II, and III we write more equations in the desired format of the extended abstract theory of composites.
  We focus on a multitude of equations involving higher order derivatives. The motivation is that results and methods in the theory of
  composites then extend to these equations.
\end{abstract}
%%%%%%%%%%%%%%%%%%%%%%%%%%%%%%%%%%%%%%%%%%%%%%%%%%%%%%%%%%%%%%%%%%%%%%%% 
\section{Introduction}
\setcounter{equation}{0}
\labsect{i}
%%%%%%%%%%%%%%%%%%%%%%%%%%%%%%%%%%%%%%%%%%%%%%%%%%%%%%%%%%%%%%%%%%%%%%%%%%%%%%%%%%%%%%%%%%%%%%%%%%%%%%%%%%%
As in Parts I, II, and III \cite{Milton:2020:UPLI, Milton:2020:UPLII, Milton:2020:UPLIII}, we cast a multitude of linear science equations in
the form encountered in the extended abstract theory of composites,
\beq \BJ(\Bx)=\BL(\Bx)\BE(\Bx)-\Bs(\Bx),\quad \BGG_1\BE=\BE,\quad\BGG_1\BJ=0,
\eeq{ad1}
now concentrating on those equations that involve higher order gradients of the fields.
%where we now allow the fields and moduli to depend on $\Bx$ and possibly $t$. 
%The first equation in \eq{ad1}, i.e., the constitutive law, is typically taken to be local in spacetime if we
%allow $x_4$ to represent time), i.e. $\BJ(\Bx,t)=\BL(\Bx,t)\BE(\Bx,t)-\Bs(\Bx,t)$.
The field $\Bs(\Bx)$ is the source term, $\BGG_1$ is a projection
operator in Fourier space, while $\BL(\Bx)$ acts locally in space
and represents the material moduli. In this Part
the fields, but not $\BL(\Bx)$, may have a dependence on time $t=x_4$.
So $\Bx$ could just represent a spatial coordinate
$\Bx=(x_1,x_2,x_3)$ or it could represent a space time coordinate
$\Bx=(x_1,x_2,x_3,x_4)$ with
$t=x_4$ representing time. Going beyond the static, quasistatic, time harmonic, and dynamic equations that we
had expressed in this form in Parts  I, II, and III our focus is on equations that involve higher order gradients of the fields.

Given any two fields $\BP_1(\Bx)$ and $\BP_2(\Bx)$ in this space of fields, we define the inner product of them
to be
\beq (\BP_1,\BP_2)=\int_{\mathbb{R}^3}(\BP_1(\Bx),\BP_2(\Bx))_{\CT}\,d\Bx,
\eeq{innp}
where $(\cdot,\cdot)_{\CT}$ is a suitable inner product on the space $\CT$
such that the projection $\BGG_1$ is selfadjoint with
respect to this inner product, and thus the space $\CE$ onto which
$\BGG_1$ projects is orthogonal to the space $\CJ$ onto which
$\BGG_2=\BI-\BGG_1$ projects. As in Part III, the integral should be over
$\mathbb{R}^4$ if the fields have a dependence on time $t=x_4$. 

We will again come across examples where $\BL$ has a nontrivial null space
or has ``infinite'' entries. As mentioned in the Introduction of Part I, 
one may be able to shift $\BL(\Bx)$ and/or its inverse by appropriate
``null-$\BT$ operators'' to remove these degeneracies or singularities.

We will not repeat the correspondence, given in the Introduction of Part I,
between the formulation given here and the
more standard  formulation, with the derivatives
of potentials explicitly entering the equations,
and with the solution involving resolvents.  

As in the preceding Parts, to avoid taking unnecessary transposes, we let $\Div$ or $\underline{\nabla}\cdot$ act on the first index of a field, and the action of $\Grad$ or $\underline{\nabla}$ produces a field, the first index of
which is associated with $\Grad$ or $\underline{\nabla}$.
%%%%%%%%%%%%%%%%%%%%%%%%%%%%%%%%%%%%%%%%%%%%%%%%%%%%%%%%%%%%%%%%%%%%%%%%%%%%%%%%%%%%%%%
\section{Electrostatics with higher order gradients}
\setcounter{equation}{0}
\labsect{ii}
%%%%%%%%%%%%%%%%%%%%%%%%%%%%%%%%%%%%%%%%%%%%%%%%%%%%%%%%%%%%%%%%%%%%%%%%%%%%%%%%%%%%%%%
The change in the net body electrical charge density is
\beq \Gr=m+\Div\Bd+\Div\Div\Bq+\Div\Div\Div\Bh+\ldots=m+\Div\Bd_T,\quad \text{where}\quad \Bd_T=\Bd-\Div\Bq+\Div\Div\Bh+\ldots,
\eeq{g.5}
where it is convenient to think of the vector field $\Bd$ as representing the density of induced electric dipoles, the second order tensor field $-\Bq$ as representing
the density of induced electric quadrupoles and the third order tensor field $\Bh$ (not, of course, to be confused with the magnetic field) as representing
the density of induced electric hexapoles, and so forth. A more precise idea of what $\Bd$, $\Bq$, and $\Bh$ are will follow soon. 
Here we use the spherically symmetric multipole expansion, see, e.g., \cite{Thompson:2004:AM}, so the multipoles
here should not be confused with the multipole expansion associated with spherical harmonics. 
The term $m$ is a source term representing the change in the density of electric monopoles.
There can also be source dipoles, hexapoles, octopoles, etc. One has a great deal of flexibility in writing the source term,
since the divergence of a source dipole density, or the double divergence of a source hexapole density, or the triple divergence of a source hexapole density
both result in the same net charge density as a source monopole density. Without loss of generality, we can assume that there is only a source $-\Bs$,
representing the change in the source dipole density,
and that $m$ (integrated) is included in it. Even this source
field is nonunique as we are free to add any divergence free field to it. Conservation of charge implies that $0=\Div\Gr=\Div(\Bd_T-\Bs)$.
%If one has elemental regions $\GO(\Bx)$ centered at the point $\Bx$, with
%volume $V(\GO(\Bx))$ and include
%all sources of induced charge $\Gr(\Bx)$ included then the induced electric multipole densities are 
%\beq \Bd(\Bx)=\frac{1}{V(\GO(\Bx))} \int_{\GO(\Bx)}\Bx'\Gr(\Bx')\,d\Bx',\quad \Bq(\Bx)=\frac{1}{V(\GO(\Bx))}\int_{\GO(\Bx)}\Bx'\otimes\Bx'\Gr(\Bx')\,d\Bx',\quad
%\Bh(\Bx)=\frac{1}{V(\GO(\Bx))}\int_{\GO(\Bx)}\Bx'\otimes\Bx'\otimes\Bx'\Gr(\Bx')\,d\Bx',
%\eeq{multm}
%and so forth. For example, the elemental regions $\GO(\Bx)$ could be small spheres centered at $\Bx$ with radius independent of $\Bx$.
%In general one expects non-local relations between these moments and the voltage field $V(\Bx)$ due to the smearing effect of integrating over $\GO(\Bx)$. Here, for consistency,
%$V(\Bx)$ should also be taken as the average of the actual voltage over $\GO(\Bx)$.

One expects there may be some nonlocal relation giving the electric potential $V$ induced by $\Bs$. An example is the screened Poisson equation
associated with Debye-H\"uckel screening \cite{Black:2018:RCM}.
As an approximation in, say, a second order gradient model,
the total displacement field $\Bd_T$ may be posited to be
related to the derivatives of the electrical potential $V$ through a constitutive law taking the form
\beq \bpm \Bd' \\ \Bq \epm=\BL\bpm
\Grad V \\ \Grad\Grad V \epm - \bpm \Bs \\ 0 \epm,
\eeq{g.6}
where $\Bd'=\Bd-\Bs$.
It follows that
\beq \BGG_1=\BV(\Bk)\equiv\frac{1}{k^2+k^4}\bpm i\Bk \\ -\Bk\otimes\Bk \epm\bpm -i\Bk & -\Bk\otimes\Bk \epm
=\frac{1}{k^2+k^4}\bpm \Bk\otimes\Bk &  -i\Bk\otimes\Bk\otimes\Bk \\ i\Bk\otimes\Bk\otimes\Bk & \Bk\otimes\Bk\otimes\Bk\otimes\Bk \epm.
\eeq{g.7}
There is no uniqueness in this formulation: for instance, suppose there is a coupling in $\BL$ between $\Bq$ and $\Grad V$, giving a contribution to $\Bq$ of
$-\BCA(\Bx)\Grad V$
where $\BCA(\Bx)$ is a third order tensor. Then the contribution of that term to the total displacement field $\Bd_T$ will be
\beq \Div(\BCA(\Bx)\Grad V)=[\Div\BCA(\Bx))]\Grad V+\BCA(\Bx)\Grad\Grad V, \eeq{g.nu}
which therefore is equivalent to the sum of two couplings: One between $\Bd$ and $\Grad V$ and the other between $\Bd$ and $\Grad\Grad V$.

A better approach, which removes this ambiguity, applies when equations are the Euler Lagrange equations arising from minimization of
the integral of some
energy $W(\Bx,\Grad V,\Grad\Grad V)-2\Bs(\Bx)\cdot\Grad V$ of the dielectric material, where $W$ is quadratic in its arguments: then expressing $W(\Grad V,\Grad\Grad V)$ as
\beq W(\Grad V,\Grad\Grad V)=\bpm \Grad V \\ \Grad\Grad V \epm\cdot\BL\bpm \Grad V \\ \Grad\Grad V \epm \eeq{hg1}
uniquely identifies $\BL$ if we require that it is selfadjoint. The fields $\Bd$ and $\Bq$ that arise in this minimization, are
then labeled as the density of electric dipoles and density of electric quadrupoles respectively. This makes precise their
definition: the constitutive law is the defining equation for  $\Bd$ and $\Bq$ with $\BL$ given by \eq{hg1}. The differential constraints
on them are a direct consequence of the Euler-Lagrange equations. This viewpoint avoids the tricky question of how $\Bd_T(\Bx)$ should be
separated it into its component fields  $\Bd$ and $\Bq$.
More generally, there could be a spatially nonlocal relation between $\Bf$ and $V$. Impressively,
Camar-Eddine and Seppecher have completely characterized  all possible  local and nonlocal possible linear electrostatic responses \cite{Camar:2002:CSD}.

In composite materials there are typically block off diagonal couplings in $\BL(\Bx)$.
                  As Lord Rayleigh observed, a uniform electric field applied to a periodic array of
                  cylinders generates an array of line force dipoles, line quadrupoles, line hexapoles, etc., positioned along the
                  cylinder axes due to the multipolar polarizability of each cylinder \cite{Rayleigh:1892:IOA}.
                  Similarly, a uniform electric field applied to a periodic array of dielectric
                  spheres effectively generates an array of force dipoles, quadrupole, hexapoles, etc., positioned at the sphere centers
                  due to the multipolar polarizability of each sphere \cite{McPhedran:1978:CLS}. (The results also extend to periodic arrays of other
                  inclusions or sets of inclusions \cite{McPhedran:1981:EOR, Milton:1981:TPA, Nicorovici:1993:TPT, Yardley:1999:AFR}, although
                  the analytic extension of the field that is outside the inclusion to the inside of the inclusion has singularities not necessarily just at one point
                  but could have branch cuts, as can be the case if the inclusion has sharp corners. Also, some composite geometries can be biconnected, like a porous rock,
                  so one cannot identify one phase as the inclusion phase). If one is interested
                  in the response of the composite to applied electric fields that are not uniform, then all the multipoles will come into play when one seeks
                  in knowing the net electrical force density $\Bf(\Bx)$, with the higher order terms becoming more important the steeper the gradient in the applied field.
                  The multipolar polarizability
                  of each inclusion is replaced by an equivalent multipolar polarizability of a line or point object at the cylinder or sphere center that generates
                  dipoles, quadrupoles, hexapoles, etc., concentrated at that line or point in accordance with the
                  values of $\Grad V_e$, $\Grad\Grad V_e$, and  $\Grad\Grad\Grad V_e$ at that line or point, where $V_e$ is the nonsingular part of the potential
                  at the line or point object (due to exterior sources including the surrounding dipoles, quadrupoles, hexapoles, etc). This is 
                  much in the same way that one has the relation \eq{g.6}. The discrete nature of these dipole, quadrupole, hexapole, etc.
                  distributions means that the response will be very much dependent on where the inclusion centers are with respect to
                  the applied fields if these have high gradients. Thus the continuum representation \eq{g.6} will only be a rough approximation for periodic
                  composites. In passing, we remark that multipolar polarizabilities can be useful in the inverse problem of
                 determining an inclusion location and shape from exterior measurements of the multipolar polarizability \cite{Ammari:2006:GPT}. General shaped inclusions,
                  or even off centered cylinders or spheres have  multipolar polarizabilties with off diagonal couplings, and the values of the gradients
                  $\Grad V_e$, $\Grad\Grad V_e$, and  $\Grad\Grad\Grad V_e$ at the point where one takes the multipolar expansion 
                  are determined by boundary measurements of $V_e$, assuming $V_e$ to satisfy $\nabla^2V_e=0$.
%%%%%%%%%%%%%%%%%%%%%%%%%%%%%%%%%%%%%%%%%%%%%%%%%%%%%%%%%%%%%%%%5         
\section{Linear elasticity with higher order gradients}
\setcounter{equation}{0}
\labsect{iii}
%%%%%%%%%%%%%%%%%%%%%%%%%%%%%%%%%%%%%%%%%%%%%%%%%%%%%%%%%%55
We assume there is a density $\Bm$ of force monopoles (that acts as a source term) which might be due to say fields
permeating the material such as gravity, electrical fields, or magnetic influences, or inertial forces if one is in a frame of reference accelerating with the body.
Due to the flexibility, as in electrostatics, in writing the source term, we can assume that there is only the source force density $-\Bs$, which includes
$\Bm$ (integrated) and again is nonunique
since monopole sources can be expressed as the divergence of a dipole field in many different ways.
Balance of forces implies
\beq 0=\Div\BGs_T-\Div\Bs,\quad \BGs_T=\BGs-\Div\BM+\Div\Div\BH+\ldots,
\eeq{g.1}
where it is convenient to think of $\BGs$ as a density of induced force dipoles, $-\BM$ as a density of induced force quadrupoles poles, associated with force moments,
$\BH$ as a density of induced force hexapoles, and so forth. One expects a spatially nonlocal relation between $\Bu$ and $\Bs$. 
%Again, for consistency,
%$\Bu(\Bx)$ should also be taken as the average of the actual displacement field over $\GO(\Bx)$.
Approximating the nonlocal relation by a second order gradient theory, one may posit that
the total stress $\BGs_T$ is related to the gradients of the displacement field $\Bu$ through a constitutive law taking the form:
\beq \bpm \BGs' \\ \BM \epm=\BL\bpm
\Grad\Bu \\ \Grad\Grad\Bu \epm - \bpm \Bs \\ 0 \epm,
\eeq{g.2}
where  $\BGs'=\BGs-\Bs$,
the term $\Grad\Grad\Bu$ for example represents some sort of local bending curvature in each element in the body, and $\BM$ represents
the associated bending moments tensor. It follows that one has
$\BGG_1(\Bk)=\BV(\Bk)$, where now $\BV(\Bk)$ applied to a $(matrix, third\,\, order\,\, tensor)$ just acts on the first index of the matrix and the first two indices
of the third order tensor. 
Again there is  no uniqueness in this formulation. Ambiguities are removed when the equations are the Euler-Lagrange equations arising from minimization of the integral of some
energy $\tfrac{1}{2}W(\Bx,\Grad \Bu,\Grad\Grad \Bu)-\Bs(\Bx)\cdot\Grad \Bu$ where $W$ is quadratic in its arguments. Then $\BL$ is taken to be real and symmetric and uniquely defined
through the quadratic form associated with $W$, similar to \eq{hg1}. The fields $\BGs$ and $\BM$ are then defined by the constitutive law,
and the Euler-Lagrange equations imply the differential constraints on these fields. Again, this avoids the tricky question of
how $\BGs_T(\Bx)$ should be separated it into its component fields
$\BGs=\BGs'+\Bs$ and $\BM$. This is then easily extended to higher gradient theories. An equivalent and well known
viewpoint is to regard  $\BGs'$ and $\BM$ as Lagrange multipliers involved with minimizing the integral of $\tfrac{1}{2}W(\Bx,\BU,\BC)-\Bs\cdot\BU$ subject to the constraints that
$\Grad\Bu=\BU$ and $\Grad\Grad\Bu=\BC$. Thus one is interested in 
\beq \min_{\Bu,\BU,\BC}\int \tfrac{1}{2}W(\Bx,\BU,\BC)-\Bs'\cdot\BU-\BGs :(\Grad\Bu-\BU)-\BM\Shortstack{. . .}(\Grad\Grad\Bu-\BC)\,d\Bx,
\eeq{g.4aa}
where $:$ denotes a double contraction of indices, while $\Shortstack{. . .}$ denotes a quadruple contraction of indices.
The stationarity of the integral with respect to changes in $\Bu$, after integration by parts, yields the differential constraint that
$\Div\BGs'+\Div\Div\BM=0$, while the stationarity of the integral with respect to changes in $\BU$ and $\BC$ yields the constitutive law.
A similar perspective is adopted in thermodynamics, which has its basis in statistical physics \cite{Fisher:1964:FEM}.
 For example, the  temperature $T$ and pressure $P$ are defined as the Lagrange
multipliers associated with minimization of the energy $U(S,V)$ with respect to changes of the entropy $S$ and  changes of the volume $V$.
(One must remember that energy is conserved: this minimization is equivalent to maximization of the entropy $S(U,V)$ with
respect to changes of the energy $U$ and  changes of the volume $V$ \cite{Callen:1960:TIPa}). This interpretation as the pressure and stress as Lagrange
multipliers seems natural, but the  stress, instead of just being a Lagrange multiplier, plays a pivotal role in the source term of
Einstein's theory of general relativity.

                   In composite materials there are typically block off diagonal couplings in $\BL(\Bx)$. Thus a uniform strain applied to a periodic array of
                  cylinders effectively generates an array of line force dipoles, line quadrupoles, line hexapoles, etc., positioned along the
                  cylinder axes \cite{McPhedran:1994:RMM}. This is due to the multipolar elastic polarizability of each cylinder.
                  Similar results hold true for the propagation of elastic waves with an array of cylindrical cavities
                  \cite{Poulton:2000:EPD}.
                  If one is interested in the response of the composite to applied strains that are not uniform, then all will come into play if one is interested
                  in knowing the net force $\Bf(\Bx)$,
                  with the higher order terms becoming more important the steeper the gradient in the applied field. The discrete nature of these dipole, quadrupole, hexapole, etc.
                  distributions means that the response will be very much dependent on where the inclusion centers are with respect to
                  the applied fields if these have high gradients. So again the continuum representation \eq{g.2} will only be a rough approximation for periodic
                  composites.

                  Associated with the truncated equations is the need to include additional constraints on the fields at a boundary: see \cite{DelIsola:2015:CTA}
                  and references therein.
                  This is the usual requirement that higher order equations come with the need to fix more derivatives at the boundary to ensure uniqueness
                  of the solution.  Usually in linear elasticity one specifies the displacement $\Bu(\Bx)$ or the applied force $\Bn\cdot\BGs_0$
                  where $\Bn$ is the normal to the surface. If we consider the equations involving additionally the moments $\BM(\Bx)$
                  and the curvature term $\Grad\Grad\Bu$ then, similar to
                  plate theory, one needs to specify at the boundary the applied surface force $\Bn\cdot\BD$ and the applied surface force dipole
                  distribution $\Bn\cdot\Grad\BM$ at the surface or, alternatively, $\Bu$ and $\Grad\Bu$. In most three dimensional
                  elastic solids the fields resulting from the application of force dipoles at the surface decay rapidly away from the surface
                  (the Saint-Venant principle). On the other hand there are materials, known as second gradient elastic materials, where the decay length is very long.
                  These typically have some easy mode of deformation where certain applied uniform loadings cost little elastic energy, while gradients in the
                  average applied strain gradient cost a huge amount of elastic energy. Such materials fall outside the framework of Cauchy elasticity.

                  Although not recognized at the time, some of the first models built from a very stiff phase and a very compliant phase to have a Poisson's ratio
                  approaching $-1$ \cite{Milton:1992:CMP} are the precursors of second gradient elastic materials: see \fig{2}. I am grateful to Pierre Seppecher for this
                  insight. Three dimensional models have also been conceived \cite{Milton:2013:CCMa, Milton:2015:NET} and approximations to these
                  have been experimentally tested, at least to verify a Possion's ratio not far from $-1$ \cite{Buckmann:2014:TDD}. These models are examples
                  of affine unimode materials. Here affine means that the only macroscopic modes of easy deformation are affine ones. When the moduli are appropriately
                  scaled (so that a loading such as in \fig{2}(c) costs finite energy as opposed to the infinite energy required if the inclusions are perfectly rigid)
                  they are second gradient elastic materials. Unimode means there is only one macroscopic mode of easy deformation.
                  In theory this deformation
                  can trace any path in the space of Cauchy-Green tensors \cite{Milton:2013:CCMa}.
                  There are also examples of affine bimode materials \cite{Milton:2013:CCMb} and \fig{3}
                  shows one such example. The realization that there are practical examples of such second gradient materials, and 
                  examples of ones that have nonaffine deformations, and materials that are third gradient,
                  is due to Seppecher, Alibert, and  Dell Isola \cite{Seppecher:2011:LET}. A far grander theoretical result was obtained much earlier by
                  Camar-Eddine and Seppecher \cite{Camar:2003:DCS} who gave a complete characterization of the possible linear elastic responses, including second and third gradient
                  materials and even nonlocal responses. An important step in their analysis was the characterization of the response of arbitrary spring networks when one applies (balanced) forces at a subset of nodes, called terminal nodes, and measures the displacement as those nodes.
                  This characterization was later extended to obtain the complete characterization of the response of mass-spring networks
                  when time dependent forces are applied to the terminal nodes and the resulting time dependent displacements at those nodes
                  are measured \cite{Vasquez:2011:CCS}.
                  Related results include a complete characterization of the
                  response at  fixed frequency of electrical networks
                  of inductors,
                  capacitors, resistors and grounds \cite{Milton:2008:RRM}
 and electromagnetic cicuits \cite{Milton:2009:EC, Milton:2010:HEC},
                  and a complete characterization of the
                  forces that cable webs under tension can support \cite{Milton:2017:SFI, Bouchitte:2019:FCW}.

                  Unfortunately the microstructures Camar-Eddine and Seppecher are extremely multiscale and impossible to construct in practice.
                  Their characterization \cite{Camar:2002:CSD} of all possible and nonlocal possible linear
                  electrostatic responses shows that no dielectric material
                  can be second gradient  or third gradient in the same sense as second gradient or third gradient elastic
                  materials that we have discussed (Pierre Seppecher, private communication).

                   \begin{figure}[!ht]
  \includegraphics[width=0.7\textwidth]{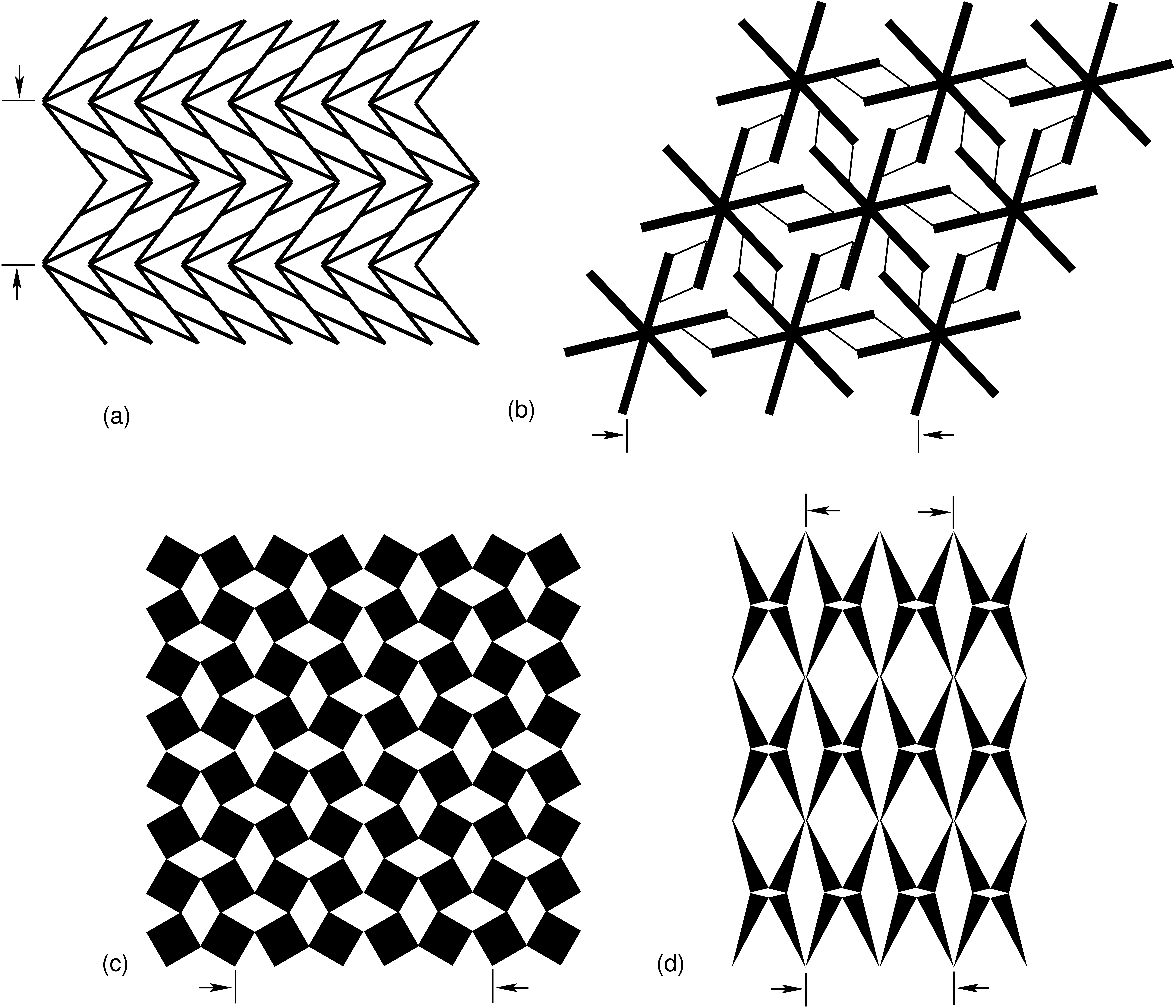}  
\caption{Four examples of affine unimode precursors to second gradient elastic materials.
  Here the black regions are rigid polygons or bars, and the points where they meet are hinge joints
  as may be theoretically achieved by gluing elastic material to them around the junction point.
  The arrows in (a), (b) and (c) show the ``pinching'' force dipoles that can uniformly shrink the entire
  material macroscopically just by a scale factor, costing little energy. These 
  In (d) such a pinching also produces a uniform deformation, however, and as is also true
  in the other examples, if, as illustrated, the pinching on the other side is opposite and the inclusions are almost rigid, then
  there will be a fight costing a lot of energy: strain gradients are penalized.
  Examples (a) and (b), drawn from {\protect\cite{Milton:1992:CMP}}
  can have a Poisson's ratio approaching $-1$. Simplified variants of these were discovered in {\protect\cite{Larsen:1997:DFC, Mitschke:2011:FAF}}
  The example (c) of Grima and Evans {\protect\cite{Grima:2000:ABR}} in closely related to an earlier one of Sigmund {\protect\cite{Sigmund:1995:TMP}}.
  The easy mode of deformation of (d) is analyzed in {\protect\cite{Milton:2013:CCMa}}. It is essentially a stacking of the pantograph structures discovered
  by Seppecher, Alibert, and Dell Isola {\protect\cite{Seppecher:2011:LET}}.
  Adapted from \protect{\cite{Milton:2013:CCMa}}, Figure 2.
}
\labfig{2}
\end{figure}

\begin{figure}[!ht]
  \includegraphics[width=0.7\textwidth]{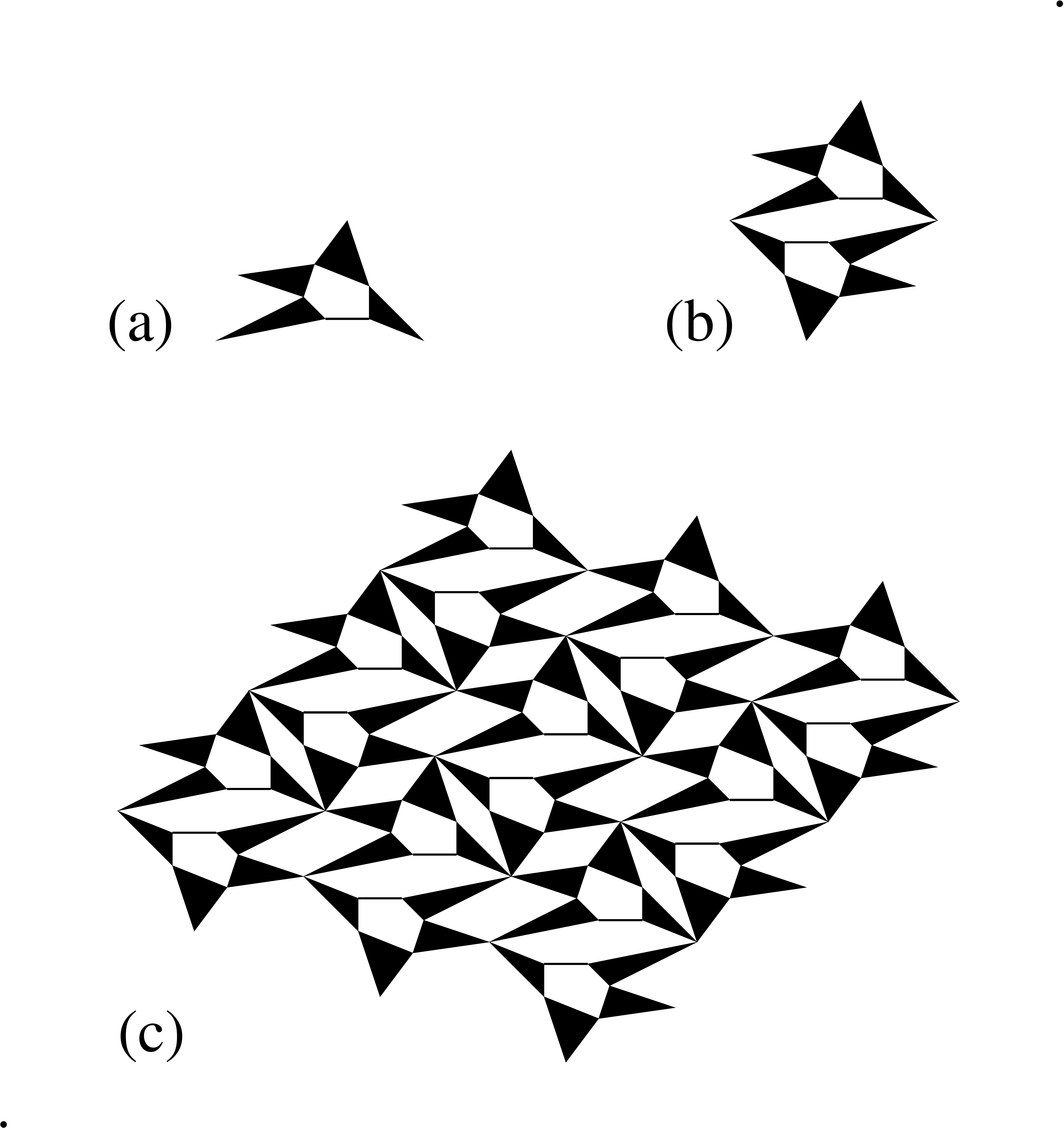}
\caption{A affine bimode precursor to a second gradient elastic material. Rigid triangles or bars are shown in black and attached by hinge joints. Figure (a) shows the basic
  underlying structure with two modes of deformation. Attaching it to a $180^\circ$ rotation of itself as in (b) gives the structure that forms the unit cell of the bimode
  material in (c). The structure in (b) has three modes of deformation, but one is incompatible with the periodicity, before deformation, of the bimode material. 
}
\labfig{3}
\end{figure}

Perhaps the best known higher order gradient equations for elasticity are the those governing the deflection and vibration of plates.
In fact, as Bigoni and Gourgiotis recognized \cite{Bigoni:2016:FFE}, a stack of stiff plates separated by say layers of rubber is a three dimensional second order
gradient metamaterial: see \fig{4}. Let us now turn to the equations describing the motions of plates and express them in the desired form.

\begin{figure}[!ht]
  \includegraphics[width=0.7\textwidth]{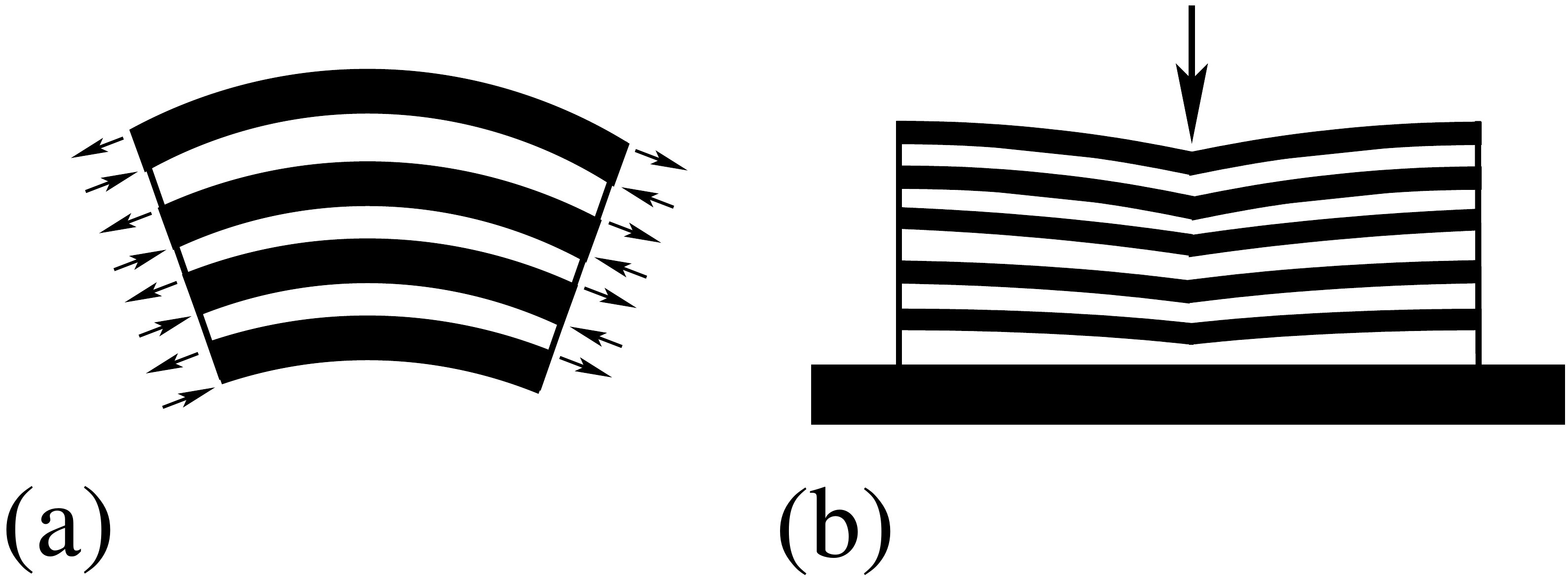}
  \caption{A stack of stiff plates separated by say layers of rubber functions as a second order gradient metamaterial \protect{\cite{Bigoni:2016:FFE}}.
      In (a) we
    see that force dipoles applied to the surface will bend the material. In (b) we see that a concentrated downward force of the stack
    will bend the plates, and accordingly require one to use the higher order gradient equations to model the deformation.
}
\labfig{4}
\end{figure}
%%%%%%%%%%%%%%%%%%%%%%%%%%%%%%%%%%%%%%%%%%%%%%%%%%%%%%%%%%%%%%%%%%%%%%%%%%
\section{Plate equations}
\setcounter{equation}{0}
\labsect{iv}
%%%%%%%%%%%%%%%%%%%%%%%%%%%%%%%%%%%%%%%%%%%%%%%%%%%%%%%%%%%%%%%%%%
%%%%%%%%%%%%%%%%%%%%%%%%%%%%%%%%%%%%%%%%%%%%%%%%%%%%%%%%%%%%%%%%%%%%%%%%%%
\subsection{Dynamic Kirchhoff–Love plate equation}
%%%%%%%%%%%%%%%%%%%%%%%%%%%%%%%%%%%%%%%%%%%%%%%%%%%%%%%%%%%%%%%%%%

With $\Bx=(x_1,x_2)$ representing coordinates in the plane of the plate (i.e. with $x_3$ being perpendicular
to the plate) these take the form \cite{Love:1888:SFV, Timoshenko:1959:TPS}.
\beq
\begin{pmatrix}
\partial \BM/\Md t \\
-\nabla\cdot(\nabla\cdot\BM)
\end{pmatrix}
= \BL \begin{pmatrix}
-\nabla\nabla v
\\
\partial v/\Md t 
\end{pmatrix} -\bpm 0 \\ \BS \epm ,
\eeq{15.15}
where $\BM(\Bx,t)$ and $v(\Bx,t)$ are the bending moment tensor and time derivative of the vertical displacement $w$ of the plate (i.e. $v=\Md w/\Md t$), and $\BS(\Bx,t)$ is a bending moment source.
Then we have
\beqa \BL(\Bx)  =  \begin{pmatrix}
-\BCD(\Bx) h(\Bx)^3 & 0 \\
0 & h(\Bx)\Gr(\Bx)
\end{pmatrix},\quad \BGG_1 & = & \frac{{\BD(i\Bk,-i\Go)}{\BD(i\Bk,-i\Go)}^\dagger}{k^4+\Go^2},\quad\text{with}\quad \BD(\Grad,\Md/\Md t)=\bpm \Grad\Grad \\ \Md/\Md t\epm, \nonum
&= & \frac{1}{k^4+\Go^2}\begin{pmatrix}
\Bk\otimes\Bk\otimes\Bk\otimes\Bk & -i\Go\Bk\otimes\Bk \\
i\Go\Bk\otimes\Bk & \Go^2
\end{pmatrix},
\eeqa{x16}
where $\Bk=(k_1, k_2)$ is the Fourier variable associated with $\Bx=(x_1,x_2)$, $h(\Bx)$ is the plate thickness, $\Gr(\Bx)$ its density, and
$\CD(\Bx)$ is the fourth order tensor of plate rigidity coefficients.

The static thin plate equations are mathematically equivalent to the static two dimensional equations of elasticity.
The vertical displacement $w$ plays the role of
the Airy stress function, whose double gradient rotated by $90^\circ$  gives the stress, and the moment tensor rotated by $90^\circ$ can be expressed as
the symmetrized gradient of a vector field that plays the role of the displacement field.
%%%%%%%%%%%%%%%%%%%%%%%%%%%%%%%%%%%%%%%%%%%%%%%%%%%%%%%%%%%%%%%%%%%%%%%%%%%%%%%%%%%%%%%%%%%%%%%%%%%
\subsection{Dynamic plate equations for vibration of moderately thick plates}
%%%%%%%%%%%%%%%%%%%%%%%%%%%%%%%%%%%%%%%%%%%%%%%%%%%%%%%%%%%%%%%%%%%%%%%%%%%%%%%%%%%%%%%%%%%%%%%%
For thicker plates Mindlin \cite{Mindlin:1951:IRI} derived a more comprehensive set of equations: see also \cite{Larsen:2009:TML}.
The constitutive relation can be written in the form
\beq  \begin{pmatrix} \Md \BM/\Md t \\  \Md \Bt/\Md t \\ \Bt-\Div\BM \\ \Div\Bt \end{pmatrix}  =  \BL
\begin{pmatrix} \Grad\dot{\BGy} \\ \dot{\BGy} - \Grad\dot{w} \\ \dot{\BGy} \\ \Md  \dot{w}/\Md t \end{pmatrix},
\eeq{Mid5}
where  $w(\Bx,t)$ is the out of plane deflection in the $x_3$ direction ;
$\BGy(\Bx,t)=(\Gy_1(\Bx,t),\Gy_2(\Bx,t))$ is the local rotation and $\dot{w}(\Bx,t)$ 
${\dot\BGy}(\Bx,t)$
their time derivatives; $\BM(\Bx,t)$ represents the bending moments,
$\Bt(\Bx,t)$ is the shear force that we will see equals  $\Bt'(\Bx,t)$,
while $\BS(\Bx,t)$ and $\Bf_t(\Bx,t)$ are bending moment and shear force sources.
The corresponding projection $\BGG_1$ is 
\beq \BGG_1(\Bk,\Go)=\bpm -\Bk & 0 \\ \Go\BI & \Bk \\ \Go\BI & 0 \\ 0 & \Go \epm
\bpm \Bk\otimes\Bk+2\Go^2\BI & \Go\Bk \\ \Go\Bk^T & k^2+\Go^2 \epm^{-1}
\bpm -\Bk^T &  \Go\BI & \Go\BI & 0 \\ 0 & \Bk^T & 0 & \Go \epm,
\eeq{Mid1}
where the matrix inverse is
\beq \bpm \Bk\otimes\Bk+2\Go^2\BI & \Go\Bk \\ \Go\Bk^T & k^2+\Go^2 \epm^{-1}=
\bpm c_1\Bk\otimes\Bk+\Go^2\BI & c_2\Bk \\ c_2\Bk^T & c_3 \epm,
\eeq{Mid1z}
with
\beq c_1=\frac{-\Go^2}{k^4+2\Go^2+2\Go^4},\quad c_2=\frac{-(k^2+2\Go^2)c_1}{\Go},\quad c_3=1-\frac{\Go k^2 c_2}{k^2+\Go^2}.
\eeq{Mid2z}
The tensor $\BL(\Bx)$ of material moduli is
\beq \BL=\begin{pmatrix} -\BCD h^3 & 0 & 0 & 0 \\
  0 & k \BGm h\BI & 0 & 0 \\
  0 & 0 &  \Gr \BI h^3/12   & 0 \\
  0 & 0 & 0 & \Gr h \epm,
\eeq{Mid6}
as given in Section 1.11 of \cite{Milton:2016:ETC} in which $\BCD(\Bx)$ is a fourth order bending rigidity tensor
depending on the local elastic moduli of the plate. 
$\Gr(\Bx)$ is the density; $h(\Bx)$ is the plate thickness; $\BGm(\Bx)$ is the shear modulus tensor; and $k$ is a shear correction factor taking the value $5/6$ for a plate.

%%%%%%%%%%%%%%%%%%%%%%%%%%%%%%%%%%%%%%%%%%%%%%%%%%%%%%%%%%%%%%% 
\section{Cosserat Elasticity}
\setcounter{equation}{0}
\labsect{v}
%%%%%%%%%%%%%%%%%%%%%%%%%%%%%%%%%%%%%%%%%%%%%%%%%%%%%%%%%%%%%
 The Cosserat brothers developed their extension of the elastodynamic equations back in 1909 \cite{Cosserat:1967:FPC}.
 A Cosserat medium is imagined to consist
 of particles, the position of whose centers determines the displacement field $\Bu$, and which can have microrotations given by a field $\BGt$. The particles
 position and rotations interact in a manner governed by the Cosserat equations.
The rotation field $\BGt$ should not be confused with the nonsymmetric part of $\Grad\Bu$. Here we use the form of the linearized Cosserat
 equations presented in \cite{Kulesh:2009:WLE}:
  \beqa
 & ~& \Grad[(\Gk+\tfrac{4}{3}\Gm)\Div\Bu]-2\Gm\Grad^2\Bu+2\Ga\Curl\BGt+\Bf=\Gr\frac{\Md^2\Bu}{\Md t^2}, \nonum
 &~& \Grad[(\widetilde{\Gk}+\tfrac{4}{3}\widetilde{\Gm})\Div\BGt]-2\widetilde{\Gm}\Grad^2\BGt+2\Ga\Curl\Bu-4\Ga\BGt+\bfm{\ell}
 =\BR\frac{\Md^2\BGt}{\Md t^2},
 \eeqa{cos0}
 where $\Gk$ and $\Gm$ are the bulk and shear elastic moduli contributing to the stress $\BGs$,
 $\widetilde{\Gk}$ and $\widetilde{\Gm}$ are analogous parameters contributing to the couple stress $\widetilde{\BGs}$, $\Ga$ is a coupling parameter,
 $\BR$ is the inertia moment density, $\Bf$ is the external body force source density, $\bfm{\ell}$ is the external torque force source density.
 Writing these equations in the desired format was challenging. The form below looks like it could be improved, but I have been unable to find anything simpler.
One has
 \beq \bpm 0 \\ -\Md\BGs/\Md t \\ \Div\BGs \\ \widetilde{\BGs} \\ -\widetilde{\Bp} \\ \Div\widetilde{\BGs} -\Md\widetilde{\Bp}/\Md t \epm
 =\BL \bpm \Grad\Bu \\ \Grad\Bv \\ \Md \Bv/\Md t \\ \Grad\BGt \\ \Md \BGt /\Md t \\ \BGt \epm
 -\bpm 0 \\ 0 \\ \Bf \\ 0 \\0 \\ \bfm{\ell} \epm,
 \eeq{cos1}
 with
 \beqa \BL & = & \bpm
 0 & 0& 0& 0& 0& 0 \\
  0 & -\BCC & 0& 0& 0 & 0\\
  0& 0 &\Gr\BI & -2\Ga\BGn & 0& 0 \\
  0& 0& 0 & \widetilde{\BCC} & 0& 0 \\
  0 &0 & 0 &  0 & -\BR & 0 \\
 -2\Ga\BGn & 0 & 0 & 0 & 0 & 4\Ga \epm, \nonum
 \BGG_1(\Bk,\Go) & = &\frac{1}{k^2+k^2\Go^2+\Go^4}
 \bpm i\Bk \\ \Go\Bk \\ -\Go^2 \\ 0 \epm\bpm -i\Bk & \Go\Bk & -\Go^2 & 0 \epm+
 \frac{1}{k^2+\Go^2+1}\bpm 0 & 0 &0  & 0 \\ 0 &0  &  0 & 0 \\ 0 &0  &  0 & 0 \\ 0 &0  &  0 & \BS(\Bk,\Go) \epm, \nonum
 \BS(\Bk,\Go)
 & = & \frac{1}{k^2+\Go^2+1}\begin{pmatrix}
\Bk\otimes\Bk & -\Go\Bk & ik \\ 
-\Go\Bk^T & \Go^2 & -i\Go \\ -ik & i\Go & 1
\end{pmatrix}.
 \eeqa{cos4}
 where the definition of $\BS(\Bk,\Go)$ coincides with that in
 Section 6.1 of Part III
 \cite{Milton:2020:UPLIII},
 and the entries of $\BGG_1(\Bk,\Go)$ are in blocks, accounting
 for the apparent mismatch in matrix dimensions with $\BL$. 
 Here $\BCC$ and $\widetilde{\BCC}$ are fourth order tensors, that for isotropic materials take the form
 \beq \BCC=3\Gk\BGL_h+2\Gm\BGL_s \quad \widetilde{\BCC}=3\widetilde{\Gk}\BGL_h+2\widetilde{\Gm}\BGL_s,
 \eeq{cos5}
 where $\BGL_h$, $\BGL_s$  project onto the rotationally invariant subspaces consisting of matrices proportional
 to the identity matrix and tracefree symmetric matrices.
 In this formulation the stress fields $\BGs$ and $\widetilde{\BGs}$ are symmetric, and
 it seems difficult to reformulate the equations in the desired form where this is not so.

One can introduce a nonsymmetric stress
 \beq \BGs_{ns}=\BGs-2\Ga\BGn\BGt=\BCC\Bu-2\Ga\BGn\BGt, \eeq{cos6}
 in terms of which $\BGt=-\BGn\BGs_{ns}/(4\Ga)$. 
 Then eliminating $\BGt$ from the second equation in \eq{cos0} would give a complicated equation involving just
 $\BGs_{ns}$, $\Bu$ and their space and time derivatives. Also the first equation in \eq{cos0} implies
 $\Div\BGs_{ns}=\Gr\Md^2\Bu/\Md t^2-\Bf$. We then have two equations involving just $\BGs_{ns}$ and $\Bu$.
 It may be the case that these could be expressed in the desired form, but such a formulation inevitably would
 be more complicated than \eq{cos1} and \eq{cos4}

%%%%%%%%%%%%%%%%%%%%%%%%%%%%%%%%%%%%%%%%%%%%%%%%%%%%%%%%%%%%%%%%%%%%%%
 \section{ Flexoelectric, flexomagnetic, and flexomagnetoelectric equations}
\setcounter{equation}{0}
\labsect{vi}
%%%%%%%%%%%%%%%%%%%%%%%%%%%%%%%%%%%%%%%%%%%%%%%%%%%%%%55
 Here one has first and second order derivatives of the electrical potential and displacement field with
 couplings between electric fields and mechanical fields. Neglecting higher order terms,
 the static equations take the form \cite{Yudin:2013:PGP, Krichen:2016:FPU}:
\beq
\bpm \Bd \\ \Bq \\ \BGs \\ \BM \epm=\BL\bpm
-\Grad V \\ -\Grad\Grad V \\ \Grad \Bu \\ \Grad\Grad\Bu \epm, \quad \text{where} \quad \begin{matrix} \Div\Bd-\Div\Div\Bq=0, \\ ~ \\ \quad \Div\BGs-\Div\Div\BM=0, \end{matrix}
\eeq{g.8}
and accordingly, $\BGG_1$ takes the form
\beq \BGG_1=\bpm \BV(\Bk) & 0 \\ 0 & \BV(\Bk)\epm,
\eeq{g.d}
where $\BV(\Bk)$ is defined by \eq{g.7} and in the second block where it is
applied to a $(matrix, third\,\, order\,\, tensor)$ it just acts on the first index of the matrix and the first two indices
of the third order tensor. 
Thus, flexing a material causes an electric field, and conversely a gradient in the electric field causes a strain. Of course there can be source
terms too. 
Flexoelectricity is important in many applications. Unlike normal piezoelectricity, there is still a coupling in $\BL$ between electomechanical coupling
even when the material is centrosymmetric. There is also the analogous {\bf flexomagnetism} \cite{Eliseev:2009:SFF}, where the equations take the same form as \eq{g.8} and \eq{g.d}
with $\psi(\Bx)$, the magnetic scalar potential, replacing the voltage $V(\Bx)$ (assuming there are no free currents so that the magnetic field $\Bh$ can be
expressed as $\Bh=-\Grad\Psi$).
In a two phase composite of a flexoelectric material and a flexomagnetic material
an electric field gradient can create a stress gradient in the flexoelectric material, thus providing a  stress gradient in the flexomagnetic material,
and generating a magnetic field. So, in another example of a product property, we get a magnetoelectric effect where gradients in the electric field
produce a magnetic field. At the macroscopic scale everything is coupled and one gets a flexomagnetoelectric material \cite{Eliseev:2011:SFF} with
\beq \bpm \Bd \\ \Bq \\ \Bb \\ \Bq^M \\ \BGs \\ \BM \epm=\BL\bpm
-\Grad V \\ -\Grad\Grad V \\ -\Grad \psi \\ -\Grad\Grad \psi \\\Grad \Bu \\ \Grad\Grad\Bu \epm, \quad \text{where} \quad \begin{matrix} \Div\Bd+\Div\Div\Bq=0,\\ ~ \\\Div\Bb+\Div\Div\Bq^M=0,
\\ ~  \\\Div\BGs+\Div\Div\BM=0, \end{matrix}
\eeq{g.8a}
in which $\Bq^M$ is a density of magnetic quadrupoles. Interestingly, one can also get flexoantiferromagnetism,
where strain gradients cause a gradient of the antiferromagnetic moment and hence generate a net magnetic field \cite{Kabychenkov:2019:FFE}.
We remark that pairs antiferromagnetic moments can be considered as magnetic quadrupoles.
%%%%%%%%%%%%%%%%%%%%%%%%%%%%%%%%%%%%%%%%%%%%%%%%%%%%%%%%%%%%%%%%%%%%%%%%%%%%%%%%
\section{Dynamic seepage in fissured rocks}
\setcounter{equation}{0}
\labsect{vii}
%%%%%%%%%%%%%%%%%%%%%%%%%%%%%%%%%%%%%%%%%%%%%%%%%%%%%%%%%%%%%%%%%%%%%%%%%%%%%%%%
The equations derived by Barenblatt, Zheltov, and Kochina \cite{Barenblatt:1960:BCT} for seepage in fissured rocks
are:
\beq \Gb_0\frac{\Md P}{\Md T}+\Div\Bv=0,\quad \Bv=\frac{k_1}{\Gm}\Grad P+\Gn\Gb_0\frac{\Md}{\Md t}\Grad P,
\eeq{dsf1}
where $\Bv(\Bx)$ is the velocity of seepage of the liquid, $P(\Bx)$ is the pressure distribution of the liquid in the pores,
$\Gb_0(\Bx)$ is a compressibility constant, $k_1(\Bx)$ is the porosity of the system of pores arising in Darcy's law,
$\Gm$ the dynamic viscosity of the fluid, and $\Gn(\Bx)$ is a specific characteristic of fissured rocks. 
They can be rewritten as
\beq \bpm \Bs \\ \Bv \\ R \\ -\Md \Div \Bs/\Md t+\Div\Bv+\Md R/\Md t \epm
=\BL \bpm \Md \Grad P/\Md t \\ \Grad P \\ \Md P/\Md t \\ P \epm,
\eeq{dsf2}
corresponding to the desired form with
\beq \BL=\bpm 0 & 0 & 0 & 0 \\ \Gn\Gb_0 & k_1/\Gm & 0 & 0 \\ 0 & 0 & 0 & \Gb_0 \\ 0 & 0 & 0 & 0 \epm,
\quad \BGG_1=\frac{1}{\Go^2k^2+k^2+\Go^2+1}\bpm \Go\Bk \\ i\Bk \\ -i\Go \\ 1 \epm\bpm \Go\Bk & -i\Bk & i\Go & 1 \epm,
\eeq{dsf3}
Here the constitutive law implies that $\Bs=0$, but it is introduced to correspond with the form of $\BGG_1$ (it could become nonzero if
we are able to shift $\BL$ by a null-$\BT$ operator). $R(\Bx)$ is a macroscopic liquid density relative to the density of the liquid:
the compressibility constant $\Gb_0(\Bx)$ times the pressure $P(\Bx)$.
%%%%%%%%%%%%%%%%%%%%%%%%%%%%%%%%%%%%%%%%%%%%%%%%%%%%%%%%%%%%%%%%%%%%%%%%%%%%%%%%
\section{Perturbed magnetohydrodynamic  equations in an incompressible fluid}
\setcounter{equation}{0}
\labsect{viii}
%%%%%%%%%%%%%%%%%%%%%%%%%%%%%%%%%%%%%%%%%%%%%%%%%%%%%%%%%%%%%%%%%%%%%%%%%%%%%%%%
For an incompressible conducting fluid, such as
mercury with conducting currents or inside the earth's liquid core, where the earth's magnetic field is generated,
the magnetohydrodynamic equations (see Section 2.5.2 of \cite{Kelley:2009:EIP}) are given by
\beqa D\Gr/Dt=\Md\Gr/\Md t+\Div(\Gr\Bv)& = & 0, \quad \Div\Bb=0, \nonum
\Gr D\Bv/D\Gr=\Gr[\Md \Bv/\Md t+(\Bv\cdot\Grad)\Bv]& = &-\Grad(P+b^2/2\Gm_0) +(\Bb\cdot\Grad)\Bb/\Gm_0,\nonum
\Md\Bb/\Md t& = &\Curl(\Bv\otimes\Bb)+\Curl\Bj/\Gs\Gm_0 \nonum
& = &  (\Bb\cdot\Grad)\Bv-(\Div\Bv)\Bb-(\Bv\cdot\Grad)\Bb+(\Curl\Bj)/\Gs\Gm_0,
\eeqa{mhda}
where $D/D t$ is the material derivative, $\Bv$, $\Gr$ $P$, and $\Gs$ are the fluid velocity, density, pressure, and conductivity while $\Bb$ is the magnetic field, with
$b^2=\Bb\cdot\Bb$, and $\Gm_0$ is the magnetic permeability of the vacuum. As the fluid is incompressible one wants $\Div\Bv=0$. However, we will allow for nonzero $\Div\Bv$,
letting $\Gr=\Gl_b\Div\Bv$, only obtaining  $\Div\Bv=0$ in the limit $\Gl_b\to\infty$.
%These need to be completed by the equation of state that for an incompressible liquid
%such as mercury or inside the earth's liquid core is simply $\Div\Bv=0$, while for a
%compressible fluid with high heat conductivity or for an adiabatic fluid, one takes
%\beq d(P^\Gg/\Gr^\Gg)/dt=\Md(P^\Gg/\Gr^\Gg)/\Md t+(\Bv\cdot\Grad)P^\Gg/\Gr^\Gg=
%\eeq{mhdb}
%where $\Gg=1$ for a compressible fluid with high heat conductivity.

As these are nonlinear equations we replace $\Bv(\Bx,t)$, $\Gr(\Bx,t)$, $P(\Bx,t)$, and $\Bb(\Bx,t)$, with
$\Bv(\Bx)+\Ge\Bv'(\Bx,t)$, $\Gr(\Bx)+\Ge\Gr'(\Bx,t)$, $P(\Bx)+\Ge P'(\Bx,t)$, and $\Bb(\Bx)+\Ge\Bb'(\Bx,t)$.
Then to first order in $\Ge$ the perturbations satisfy:
\beqa \Md\Gr'/\Md t + \Div(\Gr\Bv') & = & -\Div(\Gr'\Bv)=-\Bv\cdot\Grad\Gr'-(\Div\Bv)\Gr',\quad  \Div\Bb=0, \nonum
\Grad P'& = & -\Gr[\Md \Bv'/\Md t+
\Bv^T(\Grad\Bv')+(\Grad\Bv)^T\Bv']-[\Md \Bv/\Md t+(\Bv\cdot\Grad)\Bv]\Gr' \nonum
&~& -\Grad(\Bb^T)\Bb'/\Gm_0-\Bb^T(\Grad\Bb')^T+ \Bb^T(\Grad\Bb')/\Gm_0+(\Grad\Bb)^T\Bb'/\Gm_0,\nonum
\Curl\Bj' & = & \Gs\Gm_0\left[\Md\Bb'/\Md t-\Bb^T\Grad\Bv'+(\Div\Bv)\Bb'+\Bv^T\Grad\Bb'-(\Grad\Bv)^T\Bb'+(\Grad\Bb)^T\Bp'/\Gr+\Bb(\Div\Bv')\right],\nonum
&~&
\eeqa{mhdc}
where $\Bp'=\Gr\Bv'$ and in the last equation we used $\Div\Bb=0$. Reformulating these linearized equations proved difficult. After many attempts, and
replacing $\Gr'$ on the right hand sides of these equations by $\Gl_v\Div\Bv'$, and noting that
\beq \BGn(\Bj')=\Gm_0[(\Grad\Bb')^T-\Grad\Bb'], \quad \Curl\Bj'=\Div\BGn(\Bj'), \eeq{mhdr}
where the first follows from $\Bj'=\Gm_0\Curl\Bb'$, I arrived at the
following equations:
\beq \bpm -\BGs'-\BGn(\Bj') \\ 0 \\ -\Div (\BGs'+\BGn(\Bj')) \\ 0 \\ \Gr'\BI \\ \Bp' \\ \Div\Bp'+\Md \Gr'/\Md t \epm=\BL
\bpm \Grad\Bb' \\ \Md \Bb'/\Md t \\ \Bb'\\ \Grad\Div\Bv' \\ \Grad\Bv' \\ \Md \Bv'/\Md t \\ \Bv'\epm,
\eeq{mhdf}
where the constitutive law will force the stress $\BGs'$ to be proportional to $\BI$, so we may write $\BGs'=-P'\BI$, and, after taking appropriate limits,
will force $\Div\Bb'=\Div\Bv'=0$. Thus $\BGn(\Bj')$ acts as a sort of antisymmetric stress component. Note that as
\beq  -\Div (\BGs'+\BGn(\Bj'))=\Grad P'-\Curl\Bj' \eeq{mhdf1}
we can recover $\Grad P'$ and $\Curl\Bj'$ by separating the field on the left into its divergence free and irrotational (curl-free) parts.
The differential constraints imply
\beqa \BGG_1 & = & \bpm \BS(\Bk,\Go) & 0 & 0 & 0 & 0 \\ 0 & 0 & 0 & 0 & 0 \\ 0 & 0 & 0 & 0 & 0 \\ 0 & 0 & 0 & 0 & 0 \\ 0 & 0 & 0 & 0 & 0 \epm
+\frac{1}{k^4+k^2+\Go^2+1}\bpm 0 \\ -\Bk\otimes\Bk \\ i\Bk \\ -i\Go \\ 1 \epm\bpm 0 & -\Bk\otimes\Bk & -i\Bk & i\Go & 1 \epm, \nonum
\BS(\Bk,\Go) & = & \frac{1}{1+k^2+\Go^2}\begin{pmatrix}
\Bk\otimes\Bk & -\Go\Bk & i\Bk \\ 
-\Go\Bk^T & \Go^2\BI & -i\Go\BI \\
-i\Bk & i\Go\BI & \BI
\end{pmatrix},
\eeqa{mhdz}
in which the definition of $\BS(\Bk,\Go)$ coincides with that in
Section 5.3 of Part III
\cite{Milton:2020:UPLIII}
but now with the interpretation that acts on the first index of the matrix in $(matrix,vector, vector)$ fields. The matrix entering the constitutive
law is given by 
\beqa \BL & = & \bpm
\Gl_b\BGL_h+2\Gm_0\BGL_a          & 0                                        & 0        &  0    & 0      & 0    & 0\\
0                & 0                  & 0              & 0                        & 0        & 0                & 0      \\
(\Bb^T-\Bb\Tr)/\Gm_0-\Gs_0\Gm_0\Bv^T &  -\Gs_0\Gm_0\BI   & \BR     & 0        & -\Br & -\Gr\BI+ \Gs_0\Gm_0(\Grad\Bb)^T            & -\Gr(\Grad\Bv)^T  \\
0                    & 0                   & 0             & 0                        & 0        & 0                & 0      \\
0 & 0  & 0 & 0 & \Gl_v\BGL_h & 0   & 0 \\
0 & 0      & 0 & 0 & 0 & 0 & \Gr\\
0  & 0   & 0 & -\Gl_v\Bv^T & -\Gl_v(\Div\Bv)\Tr & 0   & 0 \epm, \nonum
\Br& = & \Gr\Bv^T+\Gl_v[\Md\Bv/\Md t +(\Bv\cdot\Grad)\Bv]\Tr+\Gs_0\Gm_0[\Bb^T -\Bb\Tr],\quad\BR=[\Grad \Bb)^T-\Grad \Bb]/\Gm_0-\Gs_0\Gm_0[(\Div\Bv)\BI-(\Grad\Bv)^T],\nonum
&~&
\eeqa{mhdg}
in which $\BGL_h$ and $\BGL_a$ are the projection onto the matrices proportional to $\BI$ and antisymmetric matrices, respectively, $\Tr$ denotes
the operation of taking the trace,
$\Bb^T -\Bb\Tr$ acting on a matrix $\BM$ produces the vector $\Bb^T\BM-\Bb\Tr(\BM)$, and 
$\Gl_b$ and $\Gl_v$ are large parameters that approach infinity thus forcing $\Div\Bb'=\Div\Bv'=0$, with $\Gl_v\Div\Bv'$ being replaced by $\Gr'$.
%%%%%%%%%%%%%%%%%%%%%%%%%%%%%%%%%%%%%%%%%%%%%%%%%%%%%%%%%%%%%%%%%%%%%%%%%%%%%%%%%%%%%%%%%%%%%%%%%
\section{Getting rid of superfluous fields in nonlocal wave equations}
\setcounter{equation}{0}
\labsect{ix}
%%%%%%%%%%%%%%%%%%%%%%%%%%%%%%%%%%%%%%%%%%%%%%%%%%%%%%%%%%%%%%%%%%%%%%%%%%%%%%%%%%%%%%%%%%%%%%%%%

Many wave equations, in addition to the plate equations and those of Cosserat elasticity, can have higher order gradient terms.
In this context, say for time harmonic electromagnetism, one can use the constitutive law to express $\Bh$ as a function of $\Be$ and $\Bj_f$ in two different ways.
The contribution to $\Bh$ from the ``bianisotropic'' type nonlocal coupling acting on $\Curl\Be$ can be condensed into a nonlocal operator acting on $\Be$, while
the formula relating $\Curl\Bh$ to $\Be$, $\Curl\Be$ and $\Bj_f$ can be condensed into a nonlocal operator acting on $\Be$ and $\Bj_f$ giving $\Bh$; equating
these two formulae for $\Bh$ gives a nonlocal relation between $\Be$ and $\Bj_f$, which if zero gives a nonlocal relation that $\Be$ must satisfy.

Thus
one can eliminate all electromagnetic fields  except for the electric field and free current sources. This avoids introducing the electric polarization and
magnetic polarization of the medium,
                  and the associated fields $\Bd(\Bx)$ and $\Bh(\Bx)$, and even eliminates the need for introducing $\Bb(\Bx)$.
                  As McPhedran and Melrose state in their preamble to Chapter 6 in their book \cite{Melrose:1991:EPD}, descriptions
                  based on these fields become ``cumbersome and ill defined for sufficiently general media'', and later in Section 6.3 they point out that
                  ``once one has Fourier transformed in both space and time, the separation into the electric and magnetic disturbances are ill defined'',
                  and they remark that the seemingly greater generality of formulations where $\Bd$ and $\Bh$ are nonlocally related to $\Be$ and $\Bb$ is an illusion.
                  Although their remarks were for nonlocal responses in macroscopically homogeneous media, the same conclusions carry through to  macroscopically inhomogeneous media.
                  
In this spirit, the standard and bianisotropic
                  constitutive laws can be rewritten as ones involving just $\Bh$ and $\Be$ and their gradients. There is no distinction between standard electromagnetism
                  and bianisotropic electromagnetism once higher order gradients are included. Thus phenomena in chiral media such as optical activity
                  (and the related effect in acoustics
                  called acoustical activity \cite{Pine:1970:DOA}) can equivalently
                  be ascribed to the usual bianisotropic equations (or their acoustical analog \cite{Sieck:2017:OWC}) or due to a higher order gradient effect,
                  namely where the gradient of the electric field (or double gradient of the pressure) enters the
                  constitutive relation.

                  In a similar vein,
                  the nonlocal Willis equations of elastodynamics, in the absence of ``eigenstrain terms'',
                  are equivalent to the standard time harmonic elastodynamics with nonlocal terms. The couplings can be filtered out of the equations and are unobservable
                  (in the past I have also been misguided about this). That there is
                  some nonuniqueness in his equations, leading to an equivalence class of equations, was recognized by Willis \cite{Willis:2011:ECR, Willis:2012:CER, Willis:2012:CTF} though
                  not brought to the conclusion reached here. 
                  Even nonlocal versions of the standard time harmonic elastodynamics  are far more cumbersome than necessary.
                  Gradients and divergences just get absorbed
                  into the nonlocal kernels. Even in the ensemble averaged sense considered by Willis \cite{Willis:1981:VRM, Willis:1981:VPDP, Willis:1997:DC}
                  (and which we later jointly developed further \cite{Milton:2006:MNS}) a nonlocal operator acting on
                  the ensemble averaged strain is the same as the  nonlocal operator acting on the symmetrized gradient of the ensemble averaged displacement field, and that symmetrized
                  gradient can be pulled inside the nonlocal operator. 
                  After stripping away
                  the superfluous fields, one is left with a description where a nonlocal linear operator $\BG_f$ relates the displacement field $\Bu$ to the applied body forces $\Bf$: what survives is the nonlocal linear relation  $\BG_f*\Bu=\Bf$, where $*$ denotes a convolution. To clarify this, one may consider a one-dimensional
                  Willis model:
                  \beq d\Gs/dx=dp/dt-f,\quad \Gve=du/dx, \quad \Gs=C*\Gve+S*(du/dt),\quad p=S^\dagger*\Gve+\Gr*(du/dt), \eeq{w1d}
                  where $f$ is the body force density,
                  $\Gs$, $p$, $\Gve$, and $u$, are the ensemble averaged stress, momentum, strain, and displacement, while
                  $C$, $S$, $\Gr$ are the nonlocal elasticity, coupling, and density operators, $S^\dagger$ being the adjoint of $S$, and $*$ denotes a convolution in $x$,
                  Taking Fourier transforms this becomes
                 \beqa ik\widehat{\Gs}(k,\Go) & = & -i\Go \widehat{p}(k,\Go)-\widehat{f}(k,\Go),\quad \widehat{\Gve}(k,\Go)=ik\widehat{u}(k,\Go),\nonum
                 \widehat{\Gs}(k,\Go) & = & \widehat{C}(k,\Go)\widehat{\Gve}(k,\Go)-i\Go \widehat{S}(k,\Go) \widehat{u}(k,\Go),
                 \quad \widehat{p}(k,\Go)=\overline{\widehat{S}(k,\Go)}\widehat{\Gve}(k,\Go)-i\Go\widehat{\Gr}(k,\Go)\widehat{u}(k,\Go),
                 \eeqa{w1da}
                where $\overline{\widehat{S}(k,\Go)}$ is the complex conjugate of $\widehat{S}(k,\Go)$. This reduces to 
                \beq G_f*u=f,\quad\text{with}\quad
                \widehat{G}_f(k,\Go)
                =\left[k^2\widehat{C}(k,\Go)-\Go k\left(\widehat{S}(k,\Go)+\overline{\widehat{S}(k,\Go)}\right)-\Go^2\widehat{\Gr}(k,\Go)\right].
                \eeq{w1db}
                
                Looking at it the other way, although this would make the redundancy even worse,
                if one tries to keep $\Grad\Bu$ in the nonlocal equation, then why not also try to keep
                $\Grad\Grad\Bu$ as this is important in higher order gradient theory?

                We emphasize that in a nonlocal theory the stress itself is unobservable
                (except when it arises from the ensemble averaging of problems having a local constitutive law). Creating a small cavity in the
                material and measuring the tractions at the boundary of the cavity needed to keep the deformation away from the hole in its previous
                state does not work. Creating a hole itself disturbs the nonlocal kernel near the hole, and generally there
                will be no tractions that will restore the  deformation around the hole to its previous
                state --- there will still some residual deformation that the tractions cannot remove. One sees this directly
                in second order gradient elasticity theories where one has to apply moments (force dipoles),
                or further couplings for $m$-th order gradient elasticity theories, at the surface in addition to tractions if
                one wants to  restore the  deformation around the hole to its previous state \cite{DelIsola:2015:CTA}. 
                More generally,
                in nonlocal equations interfaces have to be treated carefully. Thus consider an interface between two materials,
                with and without nonlocal responses. As one approaches the interface,
                the kernel of the operator must change, as it cannot link with fields in the material with a local response.

                For ensembles of problems having a local constitutive law, like those that Willis originally considered, the
                ensemble averaged stress is an observable quantity and so the nonlocal linear relation between the ensemble averaged displacement
                field and the applied body forces can be supplemented by one  relating the ensemble averaged stress to the ensemble averaged displacement field $\Bu(\Bx)$.
                For our one dimensional model this relation becomes
                \beq  \Gs=G_\Gs*u,\quad\text{with}\quad \widehat{G}_\Gs(k,\Go)=ik\widehat{C}(k,\Go)-i\Go\widehat{S}(k,\Go). \eeq{w1dc}
                Note that one can recover the ensemble averaged momentum from the relation $\Md\Bp/\Md t=\Div\BGs +\Bf$ and from
                the ensemble averaged $\Bu$ one can obtain the ensemble averaged strain. One cannot go back and uniquely recover the
                three functions $\widehat{C}(k,\Go)$, $\widehat{S}(k,\Go)$ and $\widehat{\Gr}(k,\Go)$ from the two functions $\widehat{G}_f(k,\Go)$
                and $\widehat{G}_\Gs(k,\Go)$. One can recover $\widehat{C}(k,\Go)$ and $\widehat{\Gr}(k,\Go)$ if one assumes a zero
                coupling  $\widehat{S}(k,\Go)=0$
                as one is free to do. In three dimensions the same argument shows that an anisotropic mass-density nonlocal operator does have
                meaning (after setting the coupling to zero) but only in the context of an ensemble average of materials
                as only then does the stress have meaning. The elasticity tensor field $\widehat{\BC}(k,\Go)$ is not uniquely defined, even
                in the absence of couplings, as one can add to it a tensor field $\Ga(\Go)\BGG_1(\Bk)$ where $\BGG_1(\Bk)$ is
                the projection onto the subspace of matrices $\BA$ such that $\Bk\cdot\BA=0$.

                The only coupled elastodynamic equations it makes physical sense to keep are
                Milton-Briane-Willis equations (see (2.4) in \cite{Milton:2006:CEM}) as they are local and have
                stress-acceleration and momentum-strain couplings with a direct meaning. While they are a limiting case of the Willis equations and
                can be replaced by a higher order gradient theory, they have the advantage, along with the local
                  bianisotropic equations \cite{Serdyukov:2001:EBAM} for electromagnetism,  that is it is easier to use physical reasoning to understand unusual couplings.
                  Additionally, the continuity conditions at interfaces are the same as those in standard  elastodynamics or standard electrodynamics:
                  continuity of the displacement field $\Bu(\Bx)$ and continuity of $\Bn\cdot\BGs$, where $\BGs(\Bx)$ is the stress and $\Bn$ the normal to the surface, or continuity
                  of the tangential components of $\Be$ and $\Bh$, across  the interface. 

                  Willis has noted \cite{Willis:2011:ECR} that uniqueness of the coupling terms in the nonlocal equations
                  can be obtained if one introduces an artificial ``eigenstrain field'' $\BGe_0(\Bx)$, and takes the ensemble average of the local elastodymamic equations,
                  \beqa \BGs_\Gn(\Bx)& = &\BC_\Gn(\Bx)(\BGe_\Gn(\Bx)-\BGe_0(\Bx)),\quad \Bp_\Gn(\Bx)=\BGr_\Gn(\Bx)\Md \Bu_\Gn/\Md t, \nonum
                  \Div\BGs_\Gn+\Bf& = & \Md \Bp_\Gn/\Md t,\quad \BGe_\Gn(\Bx)=\tfrac{1}{2}[\Grad\Bu_\Gn+(\Grad\Bu_\Gn)^T],
                \eeqa{will}
                where $\Gn$ parameterizes each element in the ensemble. Then, after ensemble averaging, Willis obtains a constitutive law
                \beq \bpm \BGs \\ \Bp \epm=\bpm \BC & \BS \\ \BS^\dagger & \BGr \epm*\bpm \BGe-\BGe_0 \\ \Bu \epm, \eeq{will1}
                in which $\BGs$, $\Bp$, $\BGe$, and $\Bu$ are the ensemble averages of  $\BGs_\Gn$, $\Bp_\Gn$, $\BGe_\Gn$, and $\Bu_\Gn$, satisfying
                \beq  \Div\BGs+\Bf = \Md \Bp/\Md t, \quad \BGe(\Bx)=\tfrac{1}{2}[\Grad\Bu+(\Grad\Bu)^T], \eeq{will2}
                and $\BC$, $\BS$, $\BS^\dagger$ and $\BGr$ are the nonlocal effective operators, with $*$ in \eq{will1} denoting a spatial convolution. One can now
                recover these non-local operators from the response of $\BGs$, $\Bp$, and $\Bu$ to changes in $\Bf$ and $\BGe_0$. However,
                the assumption that the ``eigenstrain field'' $\BGe_0(\Bx)$ is uncorrelated with $\BC_\Gn(\Bx)$ is physically unrealistic. For example,
                in a multiphase composite the assumption implies that the eigenstrain is the same for all phases and additionally it would be
                difficult to control $\BGe_0(\Bx)$ in an arbitary desired manner. Thus, at the end of the day, one
                sets the artificial eigenstrain field to zero, and this obviates the need for its introduction in the first place: $\BGe_0(\Bx)$ is itself a superfluous field.
                While with artificial
                eigenstrain fields the coupling terms  do gain a unique meaning, they are not accessible to physical experiments and complicate the description
                of the ensemble averaged response. Going further in the opposite direction, one could consider force moment tensors and ``eigencurvatures'',
                and examine the couplings with these present, but of course this would unnecessarily complicate the description even more. 
                  On the other hand, a similar criticism applies to the assumption that $\Bf$ (corresponding to an eigenstress) is
                uncorrelated with the microstructure, and the Willis formulation has the appealing feature that it treats strain sources on an equal
                footing as strain sources. A parallel extension can be obtained by assuming that $\BGe_0(\Bx)$ derives from a displacement field
                $\Bu_0(\Bx)$, i.e. $\BGe_0(\Bx)=\tfrac{1}{2}[\Grad\Bu_0+(\Grad\Bu_0)^T]$, as would be the case, for example, if $\BGe_0(\Bx)$ results from an oscillatory
                thermal expansion. Then we have a nonlocal relation of the form $\BG_f*(\Bu-\Bu_0)=\Bf$, where $\BG_f$ is the same nonlocal operator as before.

                  A similar result holds true for electrostatics and elastostatics: the fields get lost in nonlocal formulations and
                  all one is left with is a nonlocal operator relating the voltage $V(\Bx)$ or displacement vector field $\Bu(\Bx)$ to the 
                  charge density source (for dielectrics), free current density (for electrical conduction), or applied body forces (for elasticity). This is the
                  setting of results of  Camar-Eddine and Seppecher \cite{Camar:2002:CSD,Camar:2003:DCS}
                  who had characterized all possible electrostatic and elastostatic nonlocal
                  responses, as  mentioned in the previous section. We remark that there are conducting metamaterials where this description does not suffice, such as
                  interpenetrating network materials where instead of one macroscopic electrical potential, there is a set of them (each on different
                  ``networks'')                  that are
                  coupled together \cite{Khruslov:1978:ABS, Briane:1998:HSW}.

%%%%%%%%%%%%%%%%%%%%%%%%%%%%%%%%%%%%%%%%%%%%%%%%%%%%%%%%%%%%%%%%%%%%%%%%%
\section*{Acknowledgements}
GWM thanks the National Science Foundation for support through grant DMS-1814854,
and Gal Shmuel for helpful comments on the manuscript, and discussions concerning the nonlocal Willis equations.
This Part like
the previous Parts, is largely based on the
books \cite{Milton:2002:TOC, Milton:2016:ETC}
and again I thank those (cited in the acknowledgements of Part I) who helped stimulate that work and who provided feedback on the drafts of those books. In particular,
Nelson Beebe is thanked for all the work he did on preparing the books for publication and for updating the associated bibtex entries. Ross McPhedran is thanked for
bringing the superfluous nature of many fields in nonlocal equations, discussed in the last section, to the attention of the author. 
%%%%%%%%%%%%%%%%%%%%%%%%%%%%%%%%%%%%%%%%%%%%%%%%%%%%%%%%%%%%%%%%%%%%%%%%%%%%%%%%%%%%%%%%%%
%\bibliographystyle{plain}
%\bibliography{/home/milton/tcbook,/home/milton/newref}
%%%%%%%%%%%%%%%%%%%%%%%%%%%%%%%%%%%%%%%%%%%%%%%%%%%%%%%%%%%%%%%%%%%%%%%%%
\ifx \bblindex \undefined \def \bblindex #1{} \fi\ifx \bbljournal \undefined
  \def \bbljournal #1{{\em #1}\index{#1@{\em #1}}} \fi\ifx \bblnumber
  \undefined \def \bblnumber #1{{\bf #1}} \fi\ifx \bblvolume \undefined \def
  \bblvolume #1{{\bf #1}} \fi\ifx \noopsort \undefined \def \noopsort #1{}
  \fi\ifx \bblindex \undefined \def \bblindex #1{} \fi\ifx \bbljournal
  \undefined \def \bbljournal #1{{\em #1}\index{#1@{\em #1}}} \fi\ifx
  \bblnumber \undefined \def \bblnumber #1{{\bf #1}} \fi\ifx \bblvolume
  \undefined \def \bblvolume #1{{\bf #1}} \fi\ifx \noopsort \undefined \def
  \noopsort #1{} \fi

\end{document}